# Quantification of summertime water ice deposition on the Martian north polar ice cap

Adrian J. Brown[*1], Wendy M. Calvin[2,], Patricio Becerra[3], Shane Byrne[3]

[1] SETI Institute, 189 Bernardo Ave, Mountain View, CA 94043, USA
[2] Geological Sciences, University of Nevada, Reno, NV, 89557, USA
[3] Lunar and Planetary Laboratory, University of Arizona, Tucson, AZ, 85721, USA

Abstract. We use observations from the Compact Reconnaissance Imaging Spectrometer for Mars (CRISM) of the north polar cap during late summer for two Martian years, to monitor the complete summer cycle of albedo and water ice grain size in order to place quantitative limits of the amount of water ice deposited in late summer.

We establish here for the first time the complete spring to summer cycle of water ice grain sizes on the north polar cap. The apparent grain sizes grow until $L_s$=132, when they appear to shrink again, until they are obscured at the end of summer by the north polar hood.

Under the assumption that the shrinking of grain sizes is due to the deposition of find grained ice, we quantify the amount of water ice deposited per Martian boreal summer, and estimate the amount of water ice that must be transported equatorward.

Interestingly, we find that the relative amount of water ice deposited in the north cap during boreal summer (0.7-7 microns) is roughly equivalent to the average amount of water ice deposited on the south polar cap during austral summer (0.6-6 microns).

Summary. We use CRISM mapping observations of the Martian north polar cap to quantify the deposition of water ice in summer. We present albedo and water ice grain size maps of the entire north polar region at four points during Mars Year 28. Using spectra collected from Mars Years 28 and 29, we find that surface water ice grain sizes increase until $L_s$=132, after which they shrink until the north polar cap obscures them at $L_s$=165.

Key Point 1. 0.7-7 microns of water ice deposited on the north polar cap in summer
Key Point 2. Water ice grain sizes grow in spring/early summer, then shrink in mid summer

---

[*] corresponding author, email: abrown@seti.org

<u>Key Point 3</u>. Similar amounts of H2O ice are deposited each summer in north and south poles

Corresponding author:
Adrian Brown
SETI Institute
189 Bernardo Ave, Mountain View, CA 94043
ph. 650 810 0223
fax. 650 968 5830
email. abrown@seti.org

Short running title: "Deposition of H2O on Martian north pole"

## 1. Introduction

The Martian water cycle is crucial to the understanding of geodynamics of the atmosphere, surface and sub-surface of the planet (Clifford, 1993). Since the Viking mission, the north polar cap has been understood as the most important source and sink of modern day Martian water ice (Farmer et al., 1976; Kieffer et al., 1976). The sublimation and deposition of water ice and dust have created an important record of layering in the North Polar Layered Deposits (NPLD) (Fishbaugh and Head, 2005). The residual ice cap on top of the NPLD is thought to be approximately one meter thick (Byrne, 2009).

It has recently been reported that the north polar cap albedo decreases due to water ice grain growth during early summer (aerocentric longitude or $L_s$=93-127, where $L_s$=90 is northern summer solstice (Langevin et al., 2005)). The presence of large grained ice at the start of summer has been interpreted to mean that the north polar residual ice cap is in an ablative state during Martian summer and is currently losing mass, although that "due to its thin nature this situation cannot persist for long" (Byrne, 2009).

However, the observations we report here indicate that the *decreasing* albedos reported by Langevin do not persist past early summer, and in fact, the near infrared albedo *increases* from $L_s$=132-167 at the same location examined by Langevin et al.. As we shall show, this behavior has now been established over multiple Martian summers.

Kieffer (1987) first suggested the possibility of water ice being deposited onto the north polar cap to brighten the albedo of the cap. Kieffer (1990) used a thermal metamorphism model and models of ice-dust mixtures to infer that the north polar cap must be composed of either 1.) "old, coarse and clean" or 2.) "young, fine and dirty" ice. In this paper, we show definitively that it is both - only a combination of both fine and coarse grained ice can explain the CRISM spectra we will show below.

Using Mariner 9 and Viking imagery, Bass et al. (2000) determined that the albedo of the polar cap brightened during late summer, after darkening in early summer. They performed radiative transfer models of albedos that they saw during late summer using fine grained deposition with Mie spheres. They could not determine the size of the $H_2O$ ice grains or the entrained dust within the cap because they only had one albedo in the visible available.

Bass et al. (2000) suggest that in order to model their Viking Orbiter imaging data results, new layers of water ice are required during late summer - with deposition depths that total at least 14 microns (for small grains) or 25 microns (for large grains) per season. This tallies well with the findings of Hart and Jakosky (Hart and Jakosky, 1986) who found that at the Viking 2 lander site, a decrease of 5% in reflectance corresponded to 10 microns of water ice deposited during one Martian summer.

Bass et al. preferred the explanation of deposition of new ice rather than creation of suncaps or cracks (Bass et al. p. 390) because those processes require old ice with lower albedo. Bass et al. also established that the albedo brightening was not due to cloud activity by showing no increase in albedo from dark adjacent areas to the bright cap.

Bass and Paige (2000) used Viking IR thermal mapper (IRTM) and MAWD measurements to determine the peak of water vapor above the north polar cap. They found that the lowest visible albedo was between $L_s$=93-103 and water vapour was not released until $L_s$=103. They found the visible albedo increased after $L_s$=103, and the temperatures were too warm for re-deposition of $CO_2$ ice. They found the center of the cap would be an area of preferred deposition (and anomalous albedo) because it was colder than the rest of the cap. They presented a model (their Fig. 11) of the ice cap as a cold trap for water vapor.

Cantor et al. (2011) used MARCI visible image data to track dust storms and changes in the albedo of the north polar cap. Cantor et al. point out that spiral dust storms intersect the cap during summer. For example, Figures 18 and 20 of Cantor et al. Showing dust deposition resulting in decreases in albedo (especially on the edge of the cap) during storms in late summer. Therefore, dust has the effect of darkening the polar cap, whereas in this study we are interested in the phenomenon of cap brightening.

## 2. Methods

We have used two instruments to make our observations of the north polar cap. The Compact Reconnaissance Imaging Spectrometer for Mars (CRISM) is a visible to near-infrared spectrometer on Mars Reconnaissance Orbiter (MRO) spacecraft that is sensitive to near infrared (NIR) light from ~0.39 to ~3.9µm and is operated by the Applied Physics Laboratory at Johns Hopkins University (Murchie et al., 2007). The MARs Color Imager (MARCI) camera is a super wide angle, fish eye lens instrument with 1024 pixels-wide CCD, also on MRO, that is operated by Malin Space Science Systems (Malin et al., 2001).

In CRISM mapping mode 10x binning is employed in the cross-track direction, consequently the mapping swathes we use have 60 pixels across, covering approximately 10.8km on the surface with a resolution of ~182m on the surface. The length of each swathe is controlled by exposure time and is variable depending on commands sent to MRO.

We produced mosaics of all the CRISM mapping data available for each two week period (equivalent to the time of an MRO planning cycle). Each mosaic is in polar stereographic projection and stretches to 75°S. Figure 1 shows the area covered by each mosaic as a MARCI image of the north polar cap.

Water ice can be mapped on the surface using spectral band fitting techniques (Brown, 2006; Brown et al., 2008b) on the near infrared mosaics and exploiting the strong 1.25μm and 1.5μm water ice absorption bands (Brown et al., 2012). The presence of a strong water ice band is also indicative of large grained water ice, and because these size grains would not have a long residence time in the thin Martian atmosphere, they cannot be confused with water ice clouds (Brown et al., 2010), therefore for this study we have made no effort to remove atmospheric effects from the CRISM data. We have, however, used MARCI to examine the state of the atmosphere during the mid-to-late summer period for these Martian years and no significant dust storms were apparent over the polar residual cap during the relevant $L_s$=132-167 period. Since Viking we have known of thin spiral clouds that appear to form near the cap edge and move south toward the equator during summer (French and Gierasch, 1979) however these are rare (Kahn, 1984) and MARCI images confirmed these clouds do not move towards the polar cap or obscure it from view during the $L_s$=132-167 period.

Throughout this paper, we will refer to the "$H_2O$ ice index" first used by Langevin et al. (2007) and adjusted for use with CRISM by Brown et al. (2010). The formula for this index is:

$$H_2O\,index = 1 - \frac{R(1.5)}{R(1.394)^{0.7}\,R(1.75)^{0.3}} \qquad (1)$$

where $R(\lambda)$ indicates the reflectance at the wavelength $\lambda$ in μm. The index is high when water ice is present and low when it is not, and it increases as larger grain water ice is present (Warren, 1982; Brown et al., 2008a).

**Atmospheric complications**. Recent studies have highlighted the role of the atmosphere in affecting the apparent albedo of the north polar cap. Calvin et al. (2014) used visible MARCI data to observe seasonal changes in the north polar cap in a period overlapping our CRISM dataset. Observations of MARCI mosaic images over the north polar cap for this study show that the early summer period is almost completely clear of clouds ($L_s$=100-120). After $L_s$=120, patchy baroclinic (vortex) clouds appear, and by $L_s$=140, there are multiple cloud systems around the cap, though apparently not over the NPRC itself. By $L_s$=150, a thin haze of cloud envelopes the entire pole out to around 60°N. Around $L_s$=160-165, the have rapidly becomes the thick cover of the north polar hood. Benson et al. (2011) (their Figure 3b and c) shows that the transition to the thick hood occurs at $L_s$=165.

The role of atmospheric contamination is particularly important because as water ice clouds form, they will masquerade for the very signal we wish to measure, of fine grained ice forming on the cap. Therefore, we have carried out an analysis described in the Appendix to estimate when the polar cap has become too thick for reliable quantification of ice deposition. Our best estimate is that the polar hood becomes obscuring in the near infrared spectral albedo at $L_s$=165. This is in agreement with the results of Benson et al., despite the fact that MCS is sensitive to longer wavelengths than CRISM. Note that the visible observations (such as those measured by MARCI) are affected more robustly by the polar hood – the visible albedo is markedly affected by $L_s$=150.

We remark that this task of surface-atmospheric separation will become far easier to address (along with many other Martian climate questions) when a suitably purposed multiwavelength LIDAR is eventually in orbit at the Red Planet (Brown et al., 2014a).

## 3. Observations

**Grain Size and Albedo Maps**. Pole-wide maps of the $H_2O$ index and albedo are presented in Figure 2 and 3. These show that the $H_2O$ index decreases over the $L_s$=132-167 period and outlines the spatial extent of this effect. It is readily apparent from the red color that changes to green in the later mosaics that the $H_2O$ index falls across the entire cap during this time.

*Depositional patterns*. As can be seen from the CRISM mosaics, the effect is confined sharply to the north polar cap, and does not extend beyond the cap edge. This argues against an atmospheric origin for this process since CRISM has been previously be shown to be capable of detecting thin Martian water ice clouds extending beyond the edge of the cap in the southern polar region (Brown et al., 2010).

**Spectra from 'Point B'**. Figure 4 shows individual spectra (no averaging has been done) from 45ºE, 85ºN (point 550, 550 on our 1000x1000 pixel mosaics). This is as close as we could get to a point examined by Langevin et al. (Langevin et al., 2005) which they termed 'Point B' at 42.5ºE, 85.2ºN. This is very close to the point called 'McMurdo' in the Calvin and Titus study (Calvin and Titus, 2008) and is examined by Fig. 7 of the Byrne et al. (2008). The two points, 'Point B' and 'McMurdo' are both on the spur connecting Planum Boreum (the main part of the cap) to Gemini Lingula (the tongue below the main body in Figure 1). Figure 2-3 mosaics support the assumption that spectra of 'Point B' and 'McMurdo' behave similarly during late summer, therefore we only examine 'Point B' closely in this paper.

**Timing of NIR albedo increase**. The spectra in Figure 4 show an increase in albedo across the NIR (1.25-2.5 µm), particularly from $L_s$=132-153, and then the albedo appears to stabilize. It is also of interest to note the albedo at around 1 µm remains steady across the late summer, as has been reported by Calvin and Titus and Byrne et al.. This is because for a water ice spectrum, the 1 µm region is less sensitive to grain size changes than the NIR (Grenfell et al., 1981, Figure 3) particularly in the presence of contaminants, which serve to control the visible albedo (e.g. on Mars). This is discussed further below in Section 5.

Our best estimate of the timing of the albedo increase comes from Figure 4b, which shows all the available NIR spectra for MY29 at Point B. The spectra decrease from $L_s$=85-111, then hold steady until $L_s$=131-137 when the spectrum amplitude begins to increase again. Observations are ended by the polar hood.

**Interannual Variations in $H_2O$ index**. In order to establish that this process is a multi-year cycle, we have extracted individual spectra from regions close to Point B and have plotted the $H_2O$ ice index for these spectra in Figure 5. Coverage in Martian year 29 is not quite as comprehensive, however we were able to establish that the $H_2O$ index behaves in a similar manner (i.e. it decreases) through the mid-to-late summer across multiple Martian years.

## 4. Modeling

We have carried out radiative transfer modeling using the model proposed by Shkuratov et al. (1999) which is a simplified 1 dimensional model that allows us to interpret the decoupling of visible and near infrared albedos in an ice pack. The spectra were taken from Point B in mid and late summer and are shown in Figure 6a and b.

The task of modeling the infrared albedo of an ice pack is challenging, so we simplified our model and set the goals as conservatively as possible.

Our modeling goal is to quantify the amount of deposition of fine grained water ice between these two points in spacetime. To that end, we created a three-component models using fine and coarse ice and dust. We used the optical constants of water ice (Warren, 1984) and palagonite (Roush et al., 1991) as a Martian dust analog. The porosity ('q' factor in Shkuratov's equations) is set to 0.3, corresponding to 30% ice/dust and 70% pores to simulate porous snowpack. In the first run, we attempted to match the MY 28 $L_s$=135.25 spectrum from Figure 4 (also reproduced in Figure 6a). We found that it was necessary to use a coarse and fine grained water ice component in order to achieve good fits to the data. We carried out a constrained iterative fit, as described in (Brown et al., 2014b).

We found a good fit to the data was achieved when the coarse grain diameters were 1350 microns (30% by volume) and fine grained component of 25 microns (60% by volume). The dust was constrained to 10% of the mixture by volume and we found this produced a grain size of 270 microns. This is summarized in Table 1.

In our second scenario, we tried to model the $L_s$=163.85 spectrum using a ceteris paribus assumption (Reutlinger et al., 2011). We simulated the deposition of fine grained water ice by constraining the coarse and fine ice component grain size, and the dust grain size and volume fraction, and let the volume fraction of the coarse and fine grained components float.

We found that a good fit was achieved for coarse grained water ice volume fractions of 20% and fine grained ice components of 70% by volume. The results are again summarized in Table 1.

| Component | Reference | Best fit grain size (microns) | Best fit concentration $L_s$=135.25 | Best fit concentration $L_s$=163.85 |
|---|---|---|---|---|
| Coarse $H_2O$ ice | Warren (1984) | 1350 | 30 | 20 |
| Fine $H_2O$ ice | Warren (1984) | 25 | 60 | 70 |
| Palagonite (soil) | Roush et al. (1991) | 270 | 10 | 10 |

Table 1 – Details of the best fit parameters for the CRISM spectrum in Figure 5. The fit was applied over the spectral range from 1.02-2.5 microns. In this range, using wavelengths from the CRISM MSP range, there are 42 bands. The porosity was constrained to be 0.3 (30% ice, 70% vacant pores).

**Quantification of deposition**. In order to estimate how much water ice has been deposited on the north polar cap between $L_s$=135 and $L_s$=164 it is necessary to make some simplifying assumptions. We assume that the 10% extra volume of coverage of the fine grained ice all corresponds to freshly deposited water ice, roughly in a porous layer on top of the pre-existing snowpack. Because this layer makes up 10% of a 70% porous snowpack, we assume this equates to 3% total additional fine grained snow. This is an effective added layer of thickness when the new ice is evenly spread. As discussed in Brown et al. (2014), we then make the minimal assumption that the photon scattering corresponds to 1 pass through the snowpack (this gives us the minimal amount of water ice deposited to explain the signature change), and convert our 25 micron grains to 0.03*25=0.75 microns of deposition. As discussed in Brown et al. (2014b), this is conservative and in water ice in the infrared light may travel through up to 10 times as much material (or greater distances in visible light (Bohren, 1983)). This would lead to maximal estimates of 7.5 microns of deposition.

**Comparison to other RT models**. We also used the Hapke radiative transfer model used in Beccara et al. (2014) to carry out a comparison to the Shkuratov model. We found qualitative agreement between the two models (e.g. adding fine grained water ice gave higher NIR albedos). We found the Hapke model required significantly less volume of fine grained ice (0.5%) required to replicate the NIR spectral signature than the Shkuratov model (10%). For consistency with our previous work, we stick with the Shkuratov model for this paper, acknowledging that both models are only approximate. We intend to investigate these differences in greater extent in future work.

We can compare our estimates of amount of water ice deposition to the amounts of water ice and vapor from previous studies. Using TES data, Smith (2004) found that a plume of water vapor of strength 40 pr-µm (e.g. his Figure 5) was present over the north polar cap from $L_s$=90-150, when it decreases markedly to around 20 pr-µm. In broad terms, this requires that roughly 20 pr-µm of water ice is either deposited on the polar cap or travels equatorward at $L_s$=150.

Bass et al. (2000) estimated 14 microns of water ice may have been deposited on the north polar cap (we disregard their coarse grain deposition figure, which we believe to be an unlikely scenario after the results discussed herein). This leaves 6 pr-µm to travel equatorward.

Our calculations suggest that a range of 0.7-7.5 microns of water ice is deposited in the $L_s$=135-164 time period. This is about half of the Bass et al. (2000) estimate and leaves 13-19 pr-µm of water vapor to travel equatorward.

## 5. Discussion

In the near infrared, when dust or soot is present, the visible albedo (<1µm) is strongly controlled by the amount and type of contaminant. In the model we have proposed, this is the case – there is a background dust amount that "locks in" the visible albedo. The NIR albedo (>1µm) has increased with decreasing grain size, therefore we have reproduced the decoupling between visible and near infrared albedos as a result of water ice grain size decreases. This decoupling effect has been known for some time (Warren and Wiscombe, 1980) and has been observed in the Antarctic (where the water ice is much cleaner than the Martian north pole and visible albedos higher, usually > 0.9) (Grenfell et al., 1994). This situation is the key to why CRISM NIR observations are so critical to determining ice grain sizes.

We shall now run through three scenarios to explain the observed spectral changes in the NIR, finishing with our preferred alternative.

**Scenario #1 Physical brake up**. One possible explanation for the spectral changes is that the large grains of water ice are being decimated in the course of the summer, causing cracks and asperities to appear, increasing the number of scattering centers in the ice (which is equivalent to decreasing the grain size). However, this process requires an established old snowpack and so is not a preferred option.

**Scenario #2 Removal of covering of coarse grained ice**. A second alternative is that fine grained water ice that was stratigraphically lower (and optically obscured) by the large grain water ice is being revealed as the large grains sublimate entirely.

**Scenario #3 Deposition of fine grained water ice**. The scenario of water vapor or fine grained water ice is being deposited on top of the large grained water ice during this period. This is consistent with MAWD (Farmer et al., 1977), TES (Smith, 2004) and CRISM (Smith et al., 2013) observations that show a decrease in north polar water vapor after $L_s$=120, which has been used to argue for water vapor deposition in a polar circulation model that rises at the cap edge and sinks in the cap interior (Haberle and Jakosky, 1990). We favor this last alternative of ice deposition due to the requirement to add a fine grained water ice component to fit the models to the spectra, however we do not consider the simple radiative transfer modeling we have done sufficient to completely rule out the other two alternatives.

**North Polar Heat Budget Implications**. Irrespective of which of the three scenarios (or even a combination of the three) turns out to be correct, the process we have quantified here has an important effect on the energy balance and state of the residual water ice cap. A higher near infrared albedo will reflect more sunlight in the 1.25-2.5µm region, causing less heating of the ice, and therefore less sublimation. The cap may even be potentially recharged with redeposited water vapor creating fine water ice grains (Haberle and Jakosky, 1990).

The deposition patterns mapped in Figure 2 and 3 demonstrate enhancements are needed to state of the art models of the north polar cap dynamics using Martian Global Circulation Models or mesoscale models (e.g (Tyler and Barnes, 2005)).

**Comparison to south polar deposition**. The scattering calculations presented herein estimate that a thin film around 0.7-7 microns deep on average is deposited in the $L_s$=135-164 period. This compares very closely to the results of deposition on the south pole during summer, where an average layer 0.6-6 microns deep is estimated (Brown et al. (2014b)). This result may be concerning, particularly due to the large amount of water vapor present over the north polar cap in comparison the south polar cap. However, mitigating against this are the relative sizes of the caps, and the temperature of the substrates.

This interesting measurement of relative equality of water ice deposition thickness on both polar caps may just be coincidental but it invites future model comparisons.

**Volatile measurements during polar night**. We have to assume that further water ice deposition is likely after $L_s$=167, which is our last observation time for MY28. After this point, as the sun drops below the horizon in the Martian arctic, water ice deposition would ideally by measured by a future LIDAR mission (Brown et al., 2014a).

**Comparison to Antarctica**. Finally, we wish to contrast the behavior of the Martian northern ice cap with the Antarctic ice sheet. Jin et al. (2008) used MODIS data at 0.64 and 1.64µm each year from 2000 to 2005 supported by field observations to map albedos and model water ice grain size across the Antarctic and found that in November (beginning of austral summer) ice grains are typically 50 microns radius and they steadily increase in radius up to 210 microns by February. No hint of an increase in the 1.64µm channel is apparent in their maps for the five year period. The key difference between the Antarctic and the Martian north pole is the temperature and increased humidity, which allows the Antarctic ice to grow by sintering throughout summer. However, as we have shown here, the Martian ice cap is struck each year by fine grained ice which increases the albedo and decreases the $H_2O$ index.

Similar results for grain size increases in the Greenland ice sheet have been reported by Nolin and collaborators (Nolin, 1998; Nolin and Stroeve, 1997). A more recent study by Lyapustin et al. (2009) used MODIS to derive grain sizes for Greenland ice during the summer of 2004, revealing a complex pattern of increasing grain sizes as melting occurred on the periphery of the cap. The derived grain size decreased after melting removed large grains. Although we do not advocate melting on Mars, a process of ablation and sublimation may be part of the eventual explanation for this intriguing, widespread and repeatable polar phenomenon.

## 6. Conclusions

This investigation has quantified the deposition of water ice on the Martian north polar cap during late summer. We have found the following:

1. Water ice absorption bands grow on the north polar cap from the end of springtime ($L_s$=85) until $L_s$=132-137. This could be due to thermal metamorphism of water ice (Eluszkiewicz, 1993), resulting in growth of grains as they age, or to removal of a coating of fine grains by wind or sublimation, revealing larger grained ice beneath.

2. After $L_s$=132-137 (see Figure 4a and b), the absorption bands depths begin to decrease, consistent with the deposition of fine grained ice.

3. Using assumptions inherent in our radiative transfer scheme, the water ice coarse grain sizes we derive have average diameters of 1350 microns. We estimate that the fine grained ice at $L_s$=135 occupies 60% by volume and has a grain size of 25 microns. By $L_s$=164, this fraction has expanded to 70% by volume at the expense of the coarse grained ice.

4. Making simple assumptions regarding the scattering properties of the ice cap, we estimate that a thin film around 0.7-7 microns deep. This compares very closely to the results of deposition on the south pole during summer, where a layer 0.6-6 microns deep is estimated (Brown et al., 2014b).

5. Using TES data, Smith et al. (2004) found a plume of water vapor of strength 40 pr-µm over the north polar cap from $L_s$=90-150, which then decreases to around 20 pr-µm. Bass et al. (2000) estimated 14 microns of water ice may have been deposited on the north polar cap leaving 6 pr-µm to travel equatorward.

The calculations presented here suggest that a range of 0.7-7.5 microns of water ice is deposited in the $L_s$=135-164 time period. This is about half of the Bass et al. (2000) estimate and leaves 13-19 pr-µm of water vapor to travel equatorward.

6. The apparent deposition pattern of water ice on the north polar cap favors the Gemina Lingulae region of the cap and invites further mesoscale modeling of this depositional process.

## 7. Acknowledgements

We thank Mike Wolff for providing his code and expertise on DISORT_multi when preparing the Appendix. We would also like to thank the entire CRISM Team, particularly the Science Operations team at JHU APL, and also the MARCI PI (Mike Malin) and his staff.

All CRISM data used in this paper is publicly available at the Planetary Data System (PDS) Geosciences node.

This investigation was funded by NASA Grants NNX13AJ73G and NNX11AN41G in the Mars Data Analysis Program administered by Mitch Schulte.

## Appendix – Polar Hood Opacities

In order to determine the polar hood NIR opacities, which are particular to the CRISM instrument and different from visible opacities available from the MARCI dataset, we use the CRISM dataset itself to constrain the timing and thickness of the north polar hood. We do this in order to rule out atmospheric interference that would undermine our quantitative findings.

In MY28, at the start of the MRO CRISM mission, a large number of full resolution observations were obtained at the Phoenix landing site at 68.22N, 234.3E in preparation for the landing in summer of MY29. As shown in Table A.1, these run from $L_s$=162-181 and cover the period when the north polar hood moves in. We then can use them to bootstrap a polar wide background optical depth for water ice.

Inspection of the CRISM observations in Table A1 shows increasing amounts of water ice for later observations. The Phoenix observations were made in MY28, same year as the CRISM images that used to create Figure 1, 2, 3, 4 (left) and the fits shown in Figure 6a and b. We calculated the $H_2O$ index (which is a proxy for the 1.5 μm band) and these are tabulated for each observation. There is no ground ice over the Phoenix site at this time, so the signature of water ice is only due to the north polar hood.

| FRT # | $L_s$ | $H_2O$ index | Qualitative Assessment | Derived NIR optical depth of $H_2O$ ice |
|-------|-------|--------------|------------------------|------------------------------------------|
| 3CAB | 162 | 0.011 | No water ice signature | ~0 |
| 3F9A | 171 | 0.038 | Patchy cloud apparent | 0.75 |
| 419C | 176 | 0.102 | 100% water ice cloud | 4.5 |
| 4395 | 181 | 0.087 | 100% water ice cloud | 3.5 |

Table A1 - CRISM FRT observations overlapping the north polar cap over the Phoenix landing site in MY28, one Mars year before landing. Each spectrum was taken from the center of each image. The landing site is located at 68.2N, 234.3E at a MOLA elevation of -4108.660 relative to the Martian aeroid.

In order to convert the $H_2O$ index into an optical thickness of atmospheric water ice, we used the DISORT_multi program which is used to realistically model the effect of water ice and dust in the Martian atmosphere on CRISM spectra (Wiseman et al., 2014; Wolff et al., 2009). We first extracted a surface albedo spectrum for the Phoenix landing site by running DISORT in inverse mode, and obtained a spectrum of roughly "flat" albedo close to 0.25 and a 3 micron water band, ubiquitous in the spectrally featureless region of Acidalia Planitia (Horgan et al., 2009).

Using this bland spectrum as a surface albedo, and employing a lambertian scattering assumption for the surface, we then generated artificial atmospheres using DISORT_multi for different water ice optical depths, at 1nm resolution. These are shown in Table A2.

| Ice optical depth | $H_2O$ index |
|-------------------|--------------|
| 0 | 0.008 |
| 0.05 | 0.017 |
| 0.1 | 0.018 |
| 0.2 | 0.021 |
| 0.25 | 0.023 |
| 0.5 | 0.027 |

| 0.75 | 0.036 |
|------|-------|
| 1.0  | 0.041 |
| 2.0  | 0.063 |
| 3.0  | 0.079 |
| 4.0  | 0.095 |
| 4.5  | 0.103 |

Table A2 – DISORT_multi runs of varying water ice optical depth and resulting $H_2O$ index. The artificial atmospheres are generated over a featureless ground spectrum based on the Phoenix landing site, using solar incidence angle = 75°, emission angle=1°, phase angle =75°.

Using the data in Table A2, we can fill out the final column in Table A1, which assigns a derived water ice optical depth for the near infrared part of the spectrum, where the 1.5 band is located and that we have relied upon in our quantitative fitting in the paper. At $L_s$=162, the atmosphere has NIR optical opacity of approximately 0.5, at $L_s$=171 this increases markedly to around 0.75, then rapidly climbs to 4.5 at $L_s$=176, then back to 3.5 for $L_s$=181. The peak at $L_s$=176 is most likely due to baroclinic cloud activity

This method does make the assumption that the north polar hood does not increase in optical thickness from the Phoenix landing site at 68°N to the polar cap at Point B at 85°N. MARCI images support this assertion with strong concentrations of baroclinic waves circling around the edges of the pole, and Figure 3c of Benson et al. (2011) also indicates that the cap is thicker at the edges and less optically thick at the north pole, making our assumptions here appear conservative.

Based on linear extrapolation between the derived optical depths shown in the final column of Table A1, (assuming a steady build up of cloud optical depth from $L_s$=162-171) and considering a NIR optical depth of greater than 0.25 to be obscuring, we find that the north polar hood is obscuring in the NIR beyond $L_s$=165.

# REFERENCES

Bass, D.S., Herkenhoff, K.E., Paige, D.A., 2000. Variability of Mars' North Polar Water Ice Cap: I. Analysis of Mariner 9 and Viking Orbiter Imaging Data. Icarus 144, 382–396.

Bass, D.S., Paige, D.A., 2000. Variability of Mars' North Polar Water Ice Cap: II. Analysis of Viking IRTM and MAWD Data. Icarus 144, 397–409.

Becerra, P., Byrne, S., Brown, A.J., 2014. Transient Bright "Halos" on the South Polar Residual Cap of Mars: Implications for Mass-Balance. Icarus. doi:10.1016/j.icarus.2014.04.050

Benson, J.L., Kass, D.M., Kleinbˆhl, A., 2011. Mars' north polar hood as observed by the Mars Climate Sounder. J. Geophys. Res. 116, E03008.

Bohren, C.F., 1983. Colors of snow, frozen waterfalls, and icebergs. J. Opt. Soc. Am. 73, 1646–1652.

Brown, A.J., 2006. Spectral Curve Fitting for Automatic Hyperspectral Data Analysis. IEEE Trans. Geosci. Remote Sens. 44, 1601–1608. doi:10.1109/TGRS.2006.870435

Brown, A.J., Byrne, S., Tornabene, L.L., Roush, T., 2008a. Louth Crater: Evolution of a layered water ice mound. Icarus 196, 433–445.

Brown, A.J., Calvin, W.M., McGuire, P.C., Murchie, S.L., 2010. Compact Reconnaissance Imaging Spectrometer for Mars (CRISM) south polar mapping: First Mars year of observations. J. Geophys. Res. 115, doi:10.1029/2009JE003333.

Brown, A.J., Calvin, W.M., Murchie, S.L., 2012. Compact Reconnaissance Imaging Spectrometer for Mars (CRISM) north polar springtime recession mapping: First 3 Mars years of observations. J. Geophys. Res. 117, E00J20.

Brown, A.J., Michaels, T.I., Byrne, S., Sun, W., Titus, T.N., Colaprete, A., Wolff, M.J., Videen, G., Grund, C.J., 2014a. The case for a modern multiwavelength, polarization-sensitive LIDAR in orbit around Mars. J. Quant. Spectrosc. Radiat. Transf. doi:10.1016/j.jqsrt.2014.10.021

Brown, A.J., Piqueux, S., Titus, T.N., 2014b. Interannual observations and quantification of summertime H2O ice deposition on the Martian CO2 ice south polar cap. Earth Planet. Sci. Lett. 406, 102–109. doi:10.1016/j.epsl.2014.08.039

Brown, A.J., Sutter, B., Dunagan, S., 2008b. The MARTE Imaging Spectrometer Experiment: Design and Analysis. Astrobiology 8, 1001–1011. doi:10.1089/ast.2007.0142

Byrne, S., 2009. The Polar Deposits of Mars. Annu. Rev. Earth Planet. Sci. 37, 535–560.

Byrne, S., Zuber, M.T., Neumann, G.A., 2008. Interannual and seasonal behavior of Martian Residual Ice-Cap Albedo. Planet. Space Sci. 56, 194–211.

Calvin, W.M., James, P.B., Cantor, B.A., Dixon, E.M., 2014. Interannual and seasonal changes in the north polar ice deposits of Mars: Observations from MY 29–31 using MARCI. Icarus.

Calvin, W.M., Titus, T.N., 2008. Summer season variability of the north residual cap of Mars as observed by the Mars Global Surveyor thermal emission spectrometer (MGS-TES). Planet. Space Sci. 56, 212–226.

Cantor, B.A., James, P.B., Calvin, W.M., 2011. MARCI and MOC observations of the atmosphere and surface cap in the north polar region of Mars. Icarus 208, 61–81.

Clifford, S.M., 1993. A Model for the Hydrologic and Climatic Behavior of Water on Mars. J. Geophys. Res.-Planets 98, 10973–11016.

Eluszkiewicz, J., 1993. On the Microphysical State of the Martian Seasonal Caps. Icarus 103, 43–48.

Farmer, C.B., Davies, D.W., Holland, A.L., D.D., L., Doms, P.E., 1977. Mars: Water vapour observations from the Viking Orbiters. J. Geophys. Res. 82, 4225–4248.

Farmer, C.B., Davies, D.W., Laporte, D.D., 1976. Mars: Northern Summer Ice Cap—Water Vapor Observations from Viking 2. Science 194, 1339–1341.

Fishbaugh, K.E., Head, J.W., 2005. Origin and characteristics of the Mars north polar basal unit and implications for polar geologic history. Icarus 174, 444–474.

French, R.G., Gierasch, P.J., 1979. The Martian Polar Vortex: Theory of Seasonal Variation and Observations of Eolian Features. J. Geophys. Res. 84, 4634–4642.

Grenfell, T.C., Perovich, D.K., Ogren, J.A., 1981. Spectral albedos of an alipne snowpack. Cold Reg. Sci. Technol. 4, 121–127.

Grenfell, T.C., Warren, S.G., Mullen, P.C., 1994. Reflection of solar radiation by the Antarctic snow surface at ultraviolet, visible, and near-infrared wavelengths. J. Geophys. Res. 99, 18669–18684.

Haberle, R.M., Jakosky, B.M., 1990. Sublimation and transport of water from the north residual polar cap on Mars. J. Geophys. Res. 95, 1423–1437.

Hart, H.M., Jakosky, B.M., 1986. Composition and stability of the condensate observed at the Viking Lander 2 site on Mars. Icarus 66, 134–142.

Horgan, B.H., Bell, J.F., III, Noe Dobrea, E.Z., Cloutis, E.A., Bailey, D.T., Craig, M.A., Roach, L.H., Mustard, J.F., 2009. Distribution of hydrated minerals in the north polar region of Mars. J. Geophys. Res. 114.

Jin, Z., Charlock, T.P., Yang, P., Xie, Y., Miller, W., 2008. Snow optical properties for different particle shapes with application to snow grain size retrieval and MODIS/CERES radiance comparison over Antarctica. Remote Sens. Environ. 112, 3563–3581.

Kahn, R., 1984. The spatial and seasonal distribution of Martian clouds and some meteorological implications. J. Geophys. Res. 89, 6671–6688.

Kieffer, H., 1987. How dirty is Mars' north polar cap, and why isn't it black? Presented at the MEVTV Workshop on Nature and Composition of Surface Units on Mars, LPI, Houston, TX, pp. 72–73.

Kieffer, H., 1990. $H_2O$ grain size and the amount of dust in Mars' residual north polar cap. J. Geophys. Res. 95, 1481–1493.

Kieffer, H.H., Chase, S.C., Martin, T.Z., Miner, E.D., Palluconi, F.D., 1976. Martian North Pole Summer Temperatures: Dirty Water Ice. Science 194, 1341–1344.

Langevin, Y., Bibring, J.-P., Montmessin, F., Forget, F., Vincendon, M., Douté, S., Poulet, F., Gondet, B., 2007. Observations of the south seasonal cap of Mars during recession in 2004–2006 by the OMEGA visible/near-infrared imaging spectrometer on board Mars Express. J. Geophys. Res. 112, 10.1029/2006JE002841.

Langevin, Y., Poulet, F., Bibring, J.-P., Schmitt, B., Doute, S., Gondet, B., 2005. Summer Evolution of the North Polar Cap of Mars as Observed by OMEGA/Mars Express. Science 307, 1581–1584.

Lyapustin, A., Tedesco, M., Wang, Y., Aoki, T., Hori, M., Kokhanovsky, A., 2009. Retrieval of snow grain size over Greenland from MODIS. Remote Sens. Environ. 113, 1976–1987.

Malin, M.C., Bell, J.F., Calvin, W., Clancy, R.T., Haberle, R.M., James, P.B., Lee, S.W., Thomas, P.C., Caplinger, M.A., 2001. Mars Color Imager (MARCI) on the Mars Climate Orbiter. J. Geophys. Res.-Planets 106, 17651–17672.

Murchie, S., Arvidson, R., Bedini, P., Beisser, K., Bibring, J.-P., Bishop, J., Boldt, J., Cavender, P., Choo, T., Clancy, R.T., Darlington, E.H., Des Marais, D., Espiritu, R., Fort, D., Green, R., Guinness, E., Hayes, J., Hash, C., Heffernan, K., Hemmler, J., Heyler, G., Humm, D., Hutcheson, J., Izenberg, N., Lee, R., Lees, J., Lohr, D., Malaret, E., T., M., McGovern, J.A., McGuire, P., Morris, R., Mustard, J., Pelkey, S., Rhodes, E., Robinson, M., Roush, T., Schaefer, E., Seagrave, G., Seelos, F., Silverglate, P., Slavney, S., Smith, M., Shyong, W.-J., Strohbehn, K., Taylor, H., Thompson, P., Tossman, B., Wirzburger, M., Wolff, M., 2007. Compact Reconnaissance Imaging Spectrometer for Mars (CRISM) on Mars Reconnaissance Orbiter (MRO). J. Geophys. Res. 112, E05S03, doi:10.1029/2006JE002682.

Nolin, A.W., 1998. Mapping the Martian polar ice caps: Applications of terrestrial optical remote sensing methods. J. Geophys. Res.-Planets 103, 25851–25864.

Nolin, A.W., Stroeve, J., 1997. The changing albedo of the Greenland ice sheet: implications for climate modeling. Ann. Glaciol. 25, 51–57.

Reutlinger, A., Schurz, G., Hüttemann, A., 2011. Ceteris Paribus Laws.

Roush, T., Pollack, J.B., Orenberg, J., 1991. Derivation of Midinfrared (5-25 micrometer) Optical Constants of Some Sillicates and Palagonite. Icarus 94, 191–208.

Shkuratov, Y., Starukhina, L., Hoffmann, H., Arnold, G., 1999. A Model of Spectral Albedo of Particulate Surfaces: Implications for Optical Properties of the Moon. Icarus 137, 235–246.

Smith, M.D., 2004. Interannual variability in TES atmospheric observations of Mars during 1999-2003. Icarus 167, 148–165.

Smith, M.D., Wolff, M.J., Clancy, R.T., Kleinböhl, A., Murchie, S.L., 2013. Vertical distribution of dust and water ice aerosols from CRISM limb-geometry observations. J. Geophys. Res. Planets 118, 321–334.

Tyler, D., Barnes, J.R., 2005. A mesoscale model study of summertime atmospheric circulations in the north polar region of Mars. J. Geophys. Res. Planets 110, E06007.

Warren, S.G., 1982. Optical properties of snow. Rev. Geophys. 20, 67–89.

Warren, S.G., 1984. Optical constants of ice from the ultraviolet to the microwave. Appl. Opt. 23, 1206–1225.

Warren, S.G., Wiscombe, W.J., 1980. A Model for the spectral albedo of snow: II: snow containing atmospheric aerosols. J. Atmospheric Sci. 37, 2734–2745.

Wiseman, S.M., Arvidson, R.E., Wolff, M.J., Smith, M.D., Seelos, F.P., Morgan, F., Murchie, S.L., Mustard, J.F., Morris, R.V., Humm, D., McGuire, P.C., 2014. Characterization of Artifacts Introduced by the Empirical Volcano-Scan Atmospheric Correction Commonly Applied to CRISM and OMEGA Near-Infrared Spectra. Icarus. doi:10.1016/j.icarus.2014.10.012

Wolff, M.J., Smith, M.D., Clancy, R.T., Arvidson, R.E., Kahre, M., Seelos, F., Murchie, S., Savijarvi, H., 2009. Wavelength Dependence of Dust Aerosol Single Scattering Albedo As Observed by CRISM. J. Geophys. Res. 114, doi:10.1029/2009JE003350.

**FIGURES AND CAPTIONS**

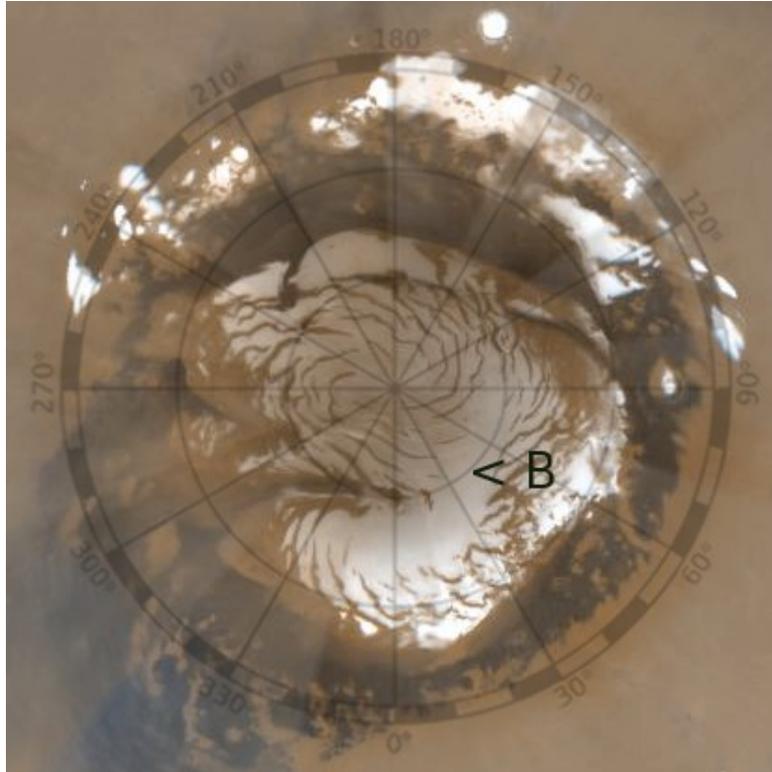

Figure 1 – MARCI image of north polar cap Mars Year 28, aerocentric longitude or $L_s$=130 (mid summer) showing location of CRISM spectra from Figure 4 (point 'B'). Outermost latitude circle is 75°S.

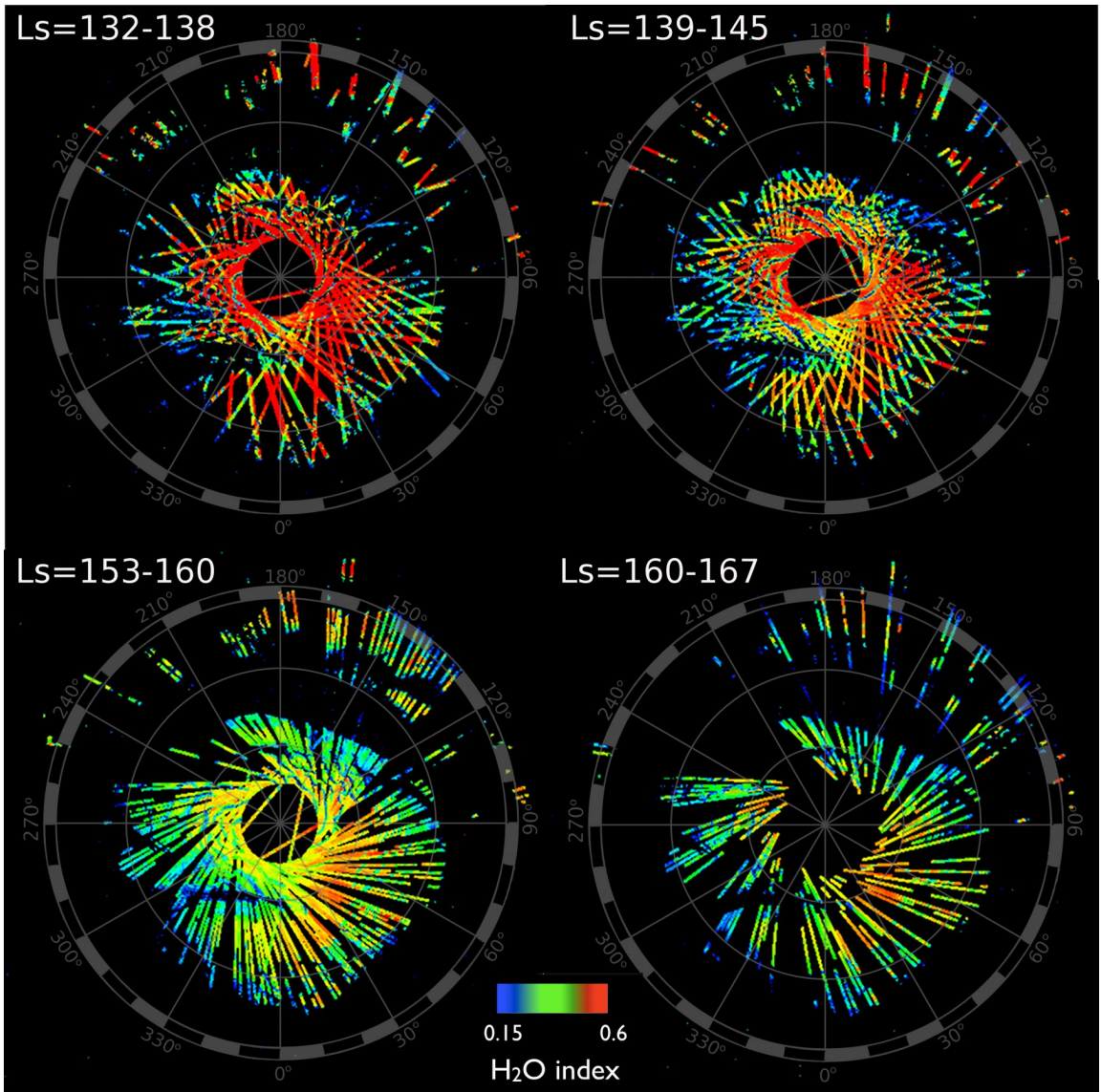

Figure 2 – Martian year 28 northern summer $H_2O$ ice index mosaics. Note high $H_2O$ index over the polar ice cap that decreases as summer progresses. Outermost latitude circle is 75°S.

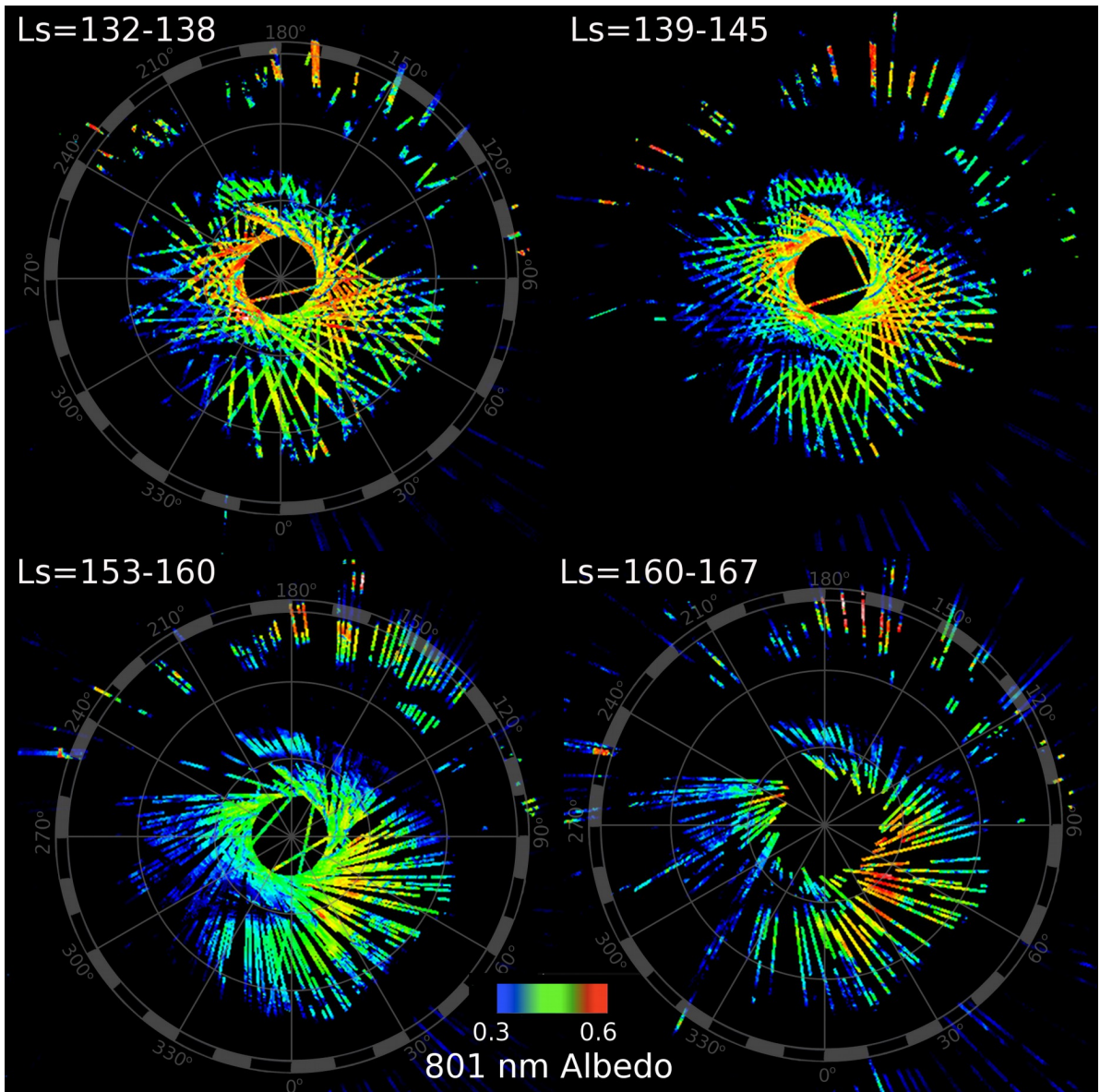

Figure 3 – Martian year 28 northern summer 801nm albedo mosaics. Note albedo is relatively steady (compared to the $H_2O$ index in Fig. 3) over the polar ice cap and slightly increases in the last panel. Outermost latitude circle is 75°S.

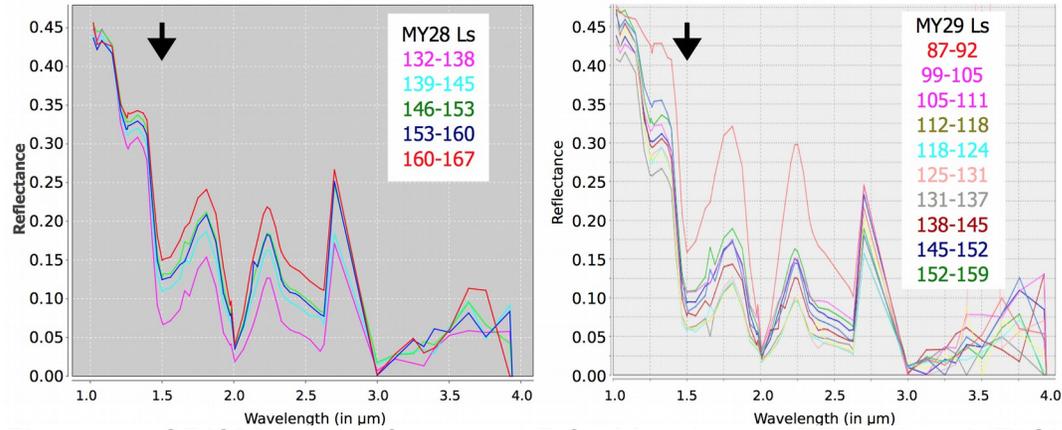

Figure 4 – CRISM spectra from point B for Mars Year (LEFT) 28 and (RIGHT) 29 in summer ($L_s$=132-167 and 87-159). All spectra from x=550,y=550 except $L_s$=160-167 which is from 550, 551 (pixels are ~180m across). Note lowest 1.5 µm albedo (largest $H_2O$ index, marked with an arrow) is achieved both years in $L_s$=132-137 period.

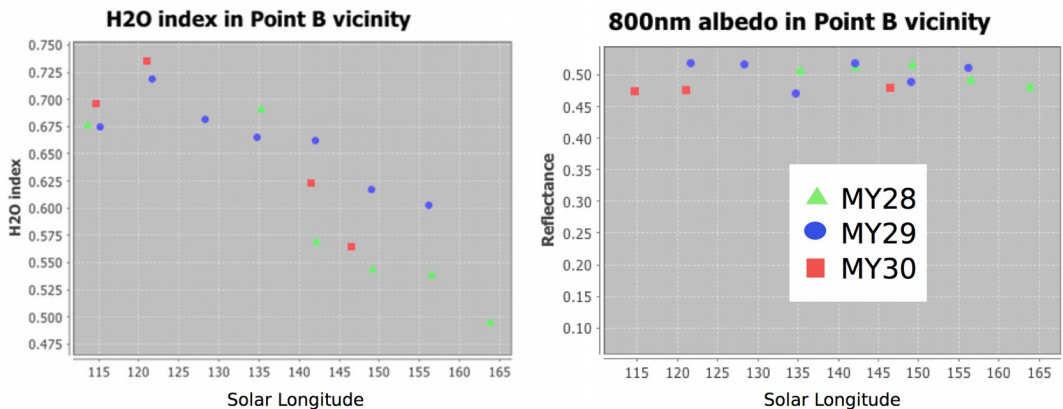

Figure 5 – (left) CRISM Mapping spectra $H_2O$ ice index taken from points close to Point B from MY 28-29 during Ls=115-160 (mid-late summer). Note decrease in $H_2O$ index from Ls=125-157 is apparent across both Mars years. (right) CRISM spectra 800nm albedo for same locations and period.

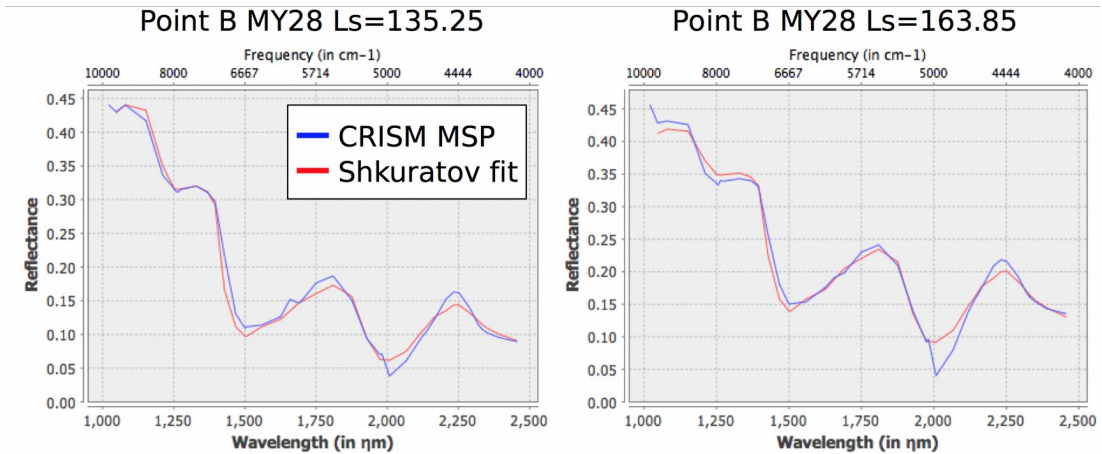

Figure 6a – (left) Point B spectrum from MY 28 $L_s$=135.25 (blue) compared with Shkuratov reflectance models of dust mixed with water ice (red). The water ice has grain sizes of 1350 (30%) and 25 (60%) microns. The dust is 10% of the mixture by volume and has a grain size of 270 microns. Figure 6b (right) As for Fig. 6a, however the season is late summer ($L_s$=163.85) and the fit includes coarse grained ice of 1350 (20%) and 25 microns (70%) resulting in a higher infrared albedo.